\documentclass[3p, 12pt, final]{elsarticle}

\biboptions{semicolon,sort&compress}

\usepackage{graphicx}                  

\biboptions{authoryear}

\usepackage[utf8]{inputenc}

\usepackage{amsfonts}
\usepackage{array}
\usepackage{amsthm}
\usepackage{amsmath}
\usepackage{bm}
\usepackage{mathptmx}
\usepackage[font={footnotesize,it},labelfont=bf]{caption}
\usepackage{subcaption}
\usepackage{hyperref}
\usepackage{booktabs}
\usepackage{multirow}
\usepackage{indentfirst}
\usepackage{amssymb,amsthm}
\usepackage{rotating}
\usepackage{pdflscape}
\usepackage{longtable}
\usepackage{xcolor}

\journal{}

\begin{document}


\begin{frontmatter}

\title{Modeling Supply-Chain Networks with Firm-to-Firm Wire Transfers\tnoteref{disclaimer}}
\tnotetext[disclaimer]{Thiago C. Silva (Grant no. 308171/2019-5, 408546/2018-2), Diego R. Amancio (Grant  no. 304026/2018-2), and Benjamin M. Tabak (Grant no. 310541/2018-2, 425123/2018-9) gratefully acknowledge financial support from the CNPq foundation. Diego R. Amancio also acknowledges financial  support from FAPESP (Grant  no. 16/19069-9)}%

 \author[ucbAddress,uspAddress]{Thiago Christiano Silva\corref{mycorrespondingauthor}}
 \ead{thiago.christiano@ucb.br}

 \author[uspAddress]{Diego Raphael Amancio}
 \ead{diego@icmc.usp.br}

 \author[fgvAddress]{Benjamin Miranda Tabak}
 \ead{benjaminm.tabak@gmail.com}

 \address[ucbAddress]{Universidade Cat\'olica de Bras\'ilia, Distrito Federal, Brazil}

\address[uspAddress]{Universidade de S\~ao Paulo, S\~ao Paulo, Brazil}

 \address[fgvAddress]{FGV/EPPG Escola de Políticas Públicas e Governo, Fundação Getúlio Vargas (School of Public Policy and Government, Getulio Vargas Foundation), Distrito Federal, Brazil}

 \cortext[mycorrespondingauthor]{Corresponding author. Address: Universidade Católica de Brasília, QS 07 – Lote 01, EPCT, CEP 71966-700, Brasília, DF.}

\begin{abstract}
We study a novel economic network (supply chain) comprised of wire transfers (electronic payment transactions) among the universe of firms in Brazil (6.2 million firms). We construct a directed and weighted network in which vertices represent cities and edges connote pairwise economic dependence between cities. Cities (vertices) represent the collection of all firms in that location and links denote intercity wire transfers. We find a high degree of economic integration among cities in the trade network, which is consistent with the high degree of specialization found across Brazilian cities. We are able to identify which cities have a dominant role in the entire supply chain process using centrality network measures. We find that the trade network has a disassortative mixing pattern, which is consistent with the power-law shape of the firm size distribution in Brazil. After the Brazilian recession in 2014, we find that the disassortativity becomes even stronger as a result of the death of many small firms and the consequent concentration of economic flows on large firms. Our results suggest that recessions have a large impact on the trade network with meaningful and heterogeneous economic consequences  across municipalities. We run econometric exercises and find that courts efficiency plays a dual role. From the customer perspective, it plays an important role in reducing contractual frictions as it increases economic transactions between different cities. From the supplier perspective, cities that are central suppliers to the supply chain seem to use courts inefficiency as a lawsuit barrier from their customers.

\end{abstract}


\end{frontmatter}

\clearpage


\section{Introduction}

Cities welfare is a crucial component to foster the development of economies both at the regional and national levels. Due to decreasing trade costs, it is notable the growing specialization of municipalities \citep{Brunelle2013}. While specialization promotes efficiency gains in the form of economies of scale, it also increases economic dependence among different cities. Such dependence creates the need of cities to transact with each other to ensure provision of all needed goods and services that are not produced locally. The economic dependence among all pairs of cities gives rise to a complex supply chain and becomes relevant to be studied under this scenario of rising specialization. Analyzing the economic flows network formed by these economic transactions can uncover many interesting aspects of cities, including their relative importance and substitutability in the national supply chain. 

The bulk of the literature on supply chains employ formal models and numerical exercises to help understand crucial aspects of supply networks (\cite{MA2020889} and \cite{CHAN2019514}). This is due to the difficulty in finding empirical databases that allow performing empirical exercises with the intent of gaining a better understanding of supply chain networks, the relevance of the nodes in the network and how they interact with other variables such as inequality, local activity and job concentration levels, city size, courts efficiency, among others.


Economic transactions among different cities are  mostly performed (in volume) by firms.  In this paper, we construct an economic flows network in which connections represent business transactions between firms from different cities. To do so, we use a novel dataset of wire electronic transfers---i.e., payments for specific goods or services from another counterparty---among any two active firms in Brazil from 2002 to 2017.  We then aggregate firms within the same municipality to construct a municipality-level network of economic flows. Therein, connections represent the sum of all business payments from two cities that arise due to firm-to-firm payments.\footnote{The more city $i$ transacts with city $j$ (by means of firms residing in each of these cities), the higher the dependence of $i$ on $j$ is. This economic dependence can arise when, for instance, an agricultural firm sells its primary goods to an industrial firm located in another city or even when it sells to an individual outside the city.} With this network, we are able to apply complex network theory to extract topological features and give economic meaning to the most widely used indicators in the related literature.

Brazil has several interesting characteristics that are worth investigating from the viewpoint of economic flows networks. First, it is an important emerging economy that is divided into five vast regions (Northeast, North, Midwest, Southeast, and South) with a total of 5,570 municipalities.  Second, due to its continental dimensions and spatial particularities, regions are subject to different climate and geographical conditions, generating the need of trading between different regions. Therefore, we expect that the number of economic transactions between firms residing in different and distant cities to be relevant. Such high number of municipalities and the growing specialization of different cities provide an ideal setup for a complex network analysis that we tackle in this paper. Third, Brazil faced a deep recession from 2014 to 2016 that had strong reflections on firm performance and probability of survival. Many small firms suffered from such recession and went bankrupt. Since our dataset goes from 2002 to 2017, we are able to understand how the recession had its tool on the structure of the economic flows network in Brazil.

Most of the works in the empirical literature that employs economic flows networks rely on very aggregate data, such as at the country level.\footnote{For instance, see \cite{Serrano2003,Fagiolo2008,Fan2014,Shen2015} for empirical research on the global economic network. A notable exception is \cite{Hussain2019}, who study a network of cities at the global level.}
When we are dealing with networks, aggregation can hide important structural aspects of the network and compromise the results and conclusions. For instance, in the case that there are many cycles in the network, the more one aggregates the network by joining vertices one into another, the fewer the number of cycles we get in the resulting network. Therefore, most of the network measures that use interconnection patterns (such as centrality indicators, network cyclicality, average geodesic path, and many others)---and therefore are sensitive to the existence of cycles---become severely compromised, normally having their levels underestimated.

Our data goes from 2003 to 2017, enabling us to understand whether the structure of the Brazilian trade network has changed over time, particularly after the global financial crisis in 2008. We find that the network has a perceptive disassortative mixing, which is consistent with the fact that the firm size distribution in Brazil has a power-law structure (few large firms and many small firms). Interestingly, we find that the network assortativity remains roughly stable from 2003 to 2014, after which it drops. This may be related to the fact that Brazil suffered a deep recession after 2014 and is being at recovery ever since. In these stressed scenario, the number of firms decreased because many firms failed, mostly likely the small ones. In this way, the network became even more disassortative, with large firms concentrating more economic transactions and hence becoming more central.

We find a large degree of economic integration among cities, which we measure using the dependence on external suppliers and customers. This high coupling corroborates the high degree of specialization of cities either in agricultural, industry, or services activities. This evidence favors David Ricardo theory, in which cities should specialize in what they enjoy comparative advantage.

By performing a network centrality analysis, we find that the São Paulo is the most central municipality in Brazil, which is consistent with its largest GDP share in Brazil. Rio de Janeiro follows. Interestingly, the centrality dynamics of São Paulo and Rio de Janeiro are strongly correlated, which reflects their strong economic relationship. While at the regional level average centralities remain roughly the same, if we dive into a state-level analysis, we find some interesting facts. For instance, in the South, Curitiba has been gaining more importance at the cost a fall in centrality of Porto Alegre. In the North region, Belem importance deeply falls after 2008 and Manaus gains the first position in that region. In the Northeast, while Recife and Salvador place at first and second in turns, we see a large increase in importance of Fortaleza, particularly after 2008.

We also run econometric exercises to understand how socioeconomic, legal and demography characteristics of Brazilian municipalities correlate with network centrality (PageRank), degree (number of suppliers and customers), and strength (total received and paid). Such analysis permits us understand the underlying drivers of these measures. We match  our city-level wire transfers data set explored in the previous section with several socioeconomic, Judiciary and demography databases. We find that courts efficiency plays a dual role. From the customer perspective, it plays an important role in reducing contractual frictions as it increases economic transactions between different cities. From the supplier perspective, cities that are central suppliers to the supply chain seem to use courts inefficiency as a lawsuit barrier from their customers.

Our approach is completely novel as this type of dataset is quite difficult to obtain and only countries in which financial authorities track these data can implement such analysis. We contribute to the literature on network analysis by constructing different indexes for each municipality in Brazil using the particularities of the economic network composed of wire transfers among cities. 


\section{Related Works}

Our work closely relates to the literature on economic flows networks. Broadly speaking, we can divide such literature into theoretical- and empirical-oriented work. While theoretical works attempt to explain the nature of connections---i.e., the underlying reasons that promote the existence or absence of links between any two economic agents (such as firms, households, cities, and countries)\footnote{For instance, \cite{Helpman2008} develop a model of heterogeneous firms to predict the existence or not of connections between different countries. They find that connections and their trade volume (link weight) yield a generalized gravity equation.}---empirical works seek to understand the role of the network structure that arise from every connection among economic agents. Our work falls in the second category.

There has been much effort in understanding economic flows networks in the empirical literature. For instance, there is a large body of research studying the world trade economic flows network. In this line of research, \cite{Serrano2003} investigate the topological characteristics of the world trade web modeled as a binary and undirected graph. The work in \citep{Fagiolo2008} instead uses a weighted network and analyzes the temporal topological properties of world trade economic flows networks. In turn, \cite{Fan2014} explores the countries' roles in the world trade network using traditional measures borrowed from the complex networks literature while the work in \citep{Shen2015} uses an approach of flow distances.  Our work contributes to this literature by providing a complex network analysis on a more disaggregate data (city level), rather than on very very aggregate data, such as the above-referenced works (country level). The use of more granular data permits us to better reflect the real nature of economic networks.

Our work also connects to the large body of literature discussing networks and financial networks: time-varying causal networks~\citep{SONG2016287}, spillovers in volatility for energy firms networks~\citep{RESTREPO2018630}, effect of networks on innovation~\citep{CHULUUN2017193}, political connections, centrality and firm innovation~\citep{TSAI2019180}, discussion of the strategic benefit of a firms centrality in its competitive advantage~\citep{LARRANETA2019}, bank-firm multiplex networks~\citep{LI2019}), risk contagion~\citep{WANG2019120842}), systemic risk measures~\citep{GUERRA2016329}, insolvency and contagion~\citep{SOUZA2015140},
directed clustering coefficients~\citep{TABAK2014211}, dynamic spanning trees~\citep{SENSOY2014387},
feedback centrality of default probabilities~\citep{SOUZA201654}, topological properties of stock market networks~\citep{TABAK20103240}, calculating systemic risk using feedback between real and financial sectors~\citep{SILVA201797},
role of financial institutions~\citep{SILVA2016130} and \citep{CAJUEIRO20086825},
financial networks and efficiency~\citep{SILVA2016247}, and estimating vulnerability and impact diffusion~\citep{SILVA2017109}.
 Our work innovates by documenting structural aspects of a city-level economic networks of an important emerging country (Brazil). Such type of data is difficult to obtain due to its secrecy.

\section{Data Description}
\label{sec:data-description}

We use transaction-level data from the SPB, which encompasses the \textit{Sistema de Transfer\^{e}ncia de Reservas} (STR) and the \textit{Sistema de Transfer\^{e}ncia de Fundos} (CIP-Sitraf), to construct our firm-to-firm network.\footnote{CIP-Sitraf clears most of the transfers in Brazil. STR is used to clear high-valued transactions. In this way, CIP-Sitraf has the largest quantity of payments, mainly of low values. STR, on the other hand, has fewer transactions but concentrate the most representative volume of monetary transfers.} The BCB maintains both STR and CIP-Sitraf, which are real-time gross settlement payment systems that record electronic interbank transactions in Brazil.

The payment transfers data comprises 6.2 million firms and has about 410 million transactions with a total commercial trading value of R\$ 48 trillion among firms between January 2003 and December 2014. To get a sense of the transacted volume, this corresponds to more than 20 times the annual nominal GDP of Brazil in 2014. Our payment data contains about 9 million firm local branches that transacted at least once.

In our analysis, we aggregate payments of all firms residing in the same city, i.e., we go from the more granular firm-time level to the less granular city-time level. The rationale is to understand the network structure of economic inflows and outflows among cities, which can reveal important insights of the economic role and function among cities. Due to this comprehensive dataset, we can map the entire network of economic dependencies among Brazilian cities over time. To avoid considering taxes and public fines, we exclude payments involving wire transfers from firms to the public administration institutions.

We classify cities as suppliers or customers by following the direction of money transfers. Suppliers are receivers of money and therefore reside in the creditor side of the monetary transaction. Customer cities are the payers of the money and are on the debtor side of the transaction. This identification permits us to navigate through the entire supply chain in Brazil. For instance, by following the chains of the supplier to the customer, we navigate downstream in the supply chain.

\section{Network measurements}

Complex networks have been used in a wide area of applications~\citep{HENRIQUE2019226}. This include areas as diverse as Biology~\citep{konini2017mean}, Natural Language Processing~\citep{akimushkin2017text,2015concentric}, Data Science~\citep{pham2015pafit}, Scientometrics~\cite{wang2017mining}, Time Series Analysis~\citep{gao2015complex}, and real-world applications~\citep{JAIN20191, BENELKOUNI2019113020, SRINIVAS2019296, NAMTIRTHA2020112859, MAJI2020113092, ZAREIE2018200}. A network comprises nodes, which are linked via edges. Several measurements have been devoted along the last years to study the network structure. Below we summarize some of the most important measures used for this purpose. These measurements are used here to study the main properties of the network modeling intercity economic flows.

\begin{enumerate}

    \item \emph{Density}: the density of measures how connected is the network. It is defined as the ration between the number of edges and the total in a complete network with the same number of nodes. In other words, if $E$ is the number of edges and $N$ is the number of nodes, the density is defined as $d = E/N(N-1)$.

    \item \emph{Degree}: the degree of a node is related to the total number of links linked to that node.

    \item \emph{Assortativity}: this measurement quantifies whether edges link nodes with similar characteristics. In this case, a network is  assortative if highly connected nodes tend to connect with other highly connected nodes. Conversely, a network is disassortative if highly connected nodes are linked with low-degree nodes. This measurement can be quantified by measuring the degree correlation of linked nodes.

    \item \emph{Diameter}: the diameter of a network is related to the concept of shortest path~\citep{Newman2003}. A shortest path is a path in a network linking two nodes with minimum length. The diameter of the network is maximum shortest path length linking the nodes in the network.

    \item \emph{PageRank}: the PageRank algorithm is a method to measure the importance of nodes in the network. Unlike other measurements based on strictly local features (such as degree) or quasi-local features (such as hierarchical measurements~\citep{ThiagoBook}), the PageRank uses global information to measure the centrality of a node. More specifically, the importance of a node is proportional to the importance of nodes linked to it. Mathematically, the PageRank $\pi(v_i)$ of node $v_i$ is computed as
    \begin{equation}
        \pi(v_i) = \frac{1-\alpha}{N} + \alpha \sum_{v_j \in V(v_i)} \frac{\pi(v_i)}{K(v_j)},
    \end{equation}
    where $\alpha$ is a constant to account for the probability that a random walker visit a randomly picked node in the whole network, $V(v_i)$ is the set of nodes connected to $v_i$ (in-going links to $v_i$) and $K(v_j)$ is the set of outgoing links from $v_j$.

\end{enumerate}


While there are many other complex network analysis, here we decided to focus on the most common measurements to characterize the main topological features of the network. A detailed description of the above measurements and more applications can be found elsewhere~\citep{ThiagoBook,newman2018networks}.

\section{Network topological analysis of the Brazilian intercity economic flows}
\label{annex:topologicalAnalysis}

In this section, we analyze the structural features of the Brazilian network of intercity economic flows using complex network theory. We follow \cite{ThiagoBook}'s classification of network measures.

\begin{figure}
    \centering
    \begin{subfigure}{.49\textwidth}
      \centering
      \includegraphics[width=\textwidth,clip=TRUE,trim=0cm 2cm 0cm 1.7cm]{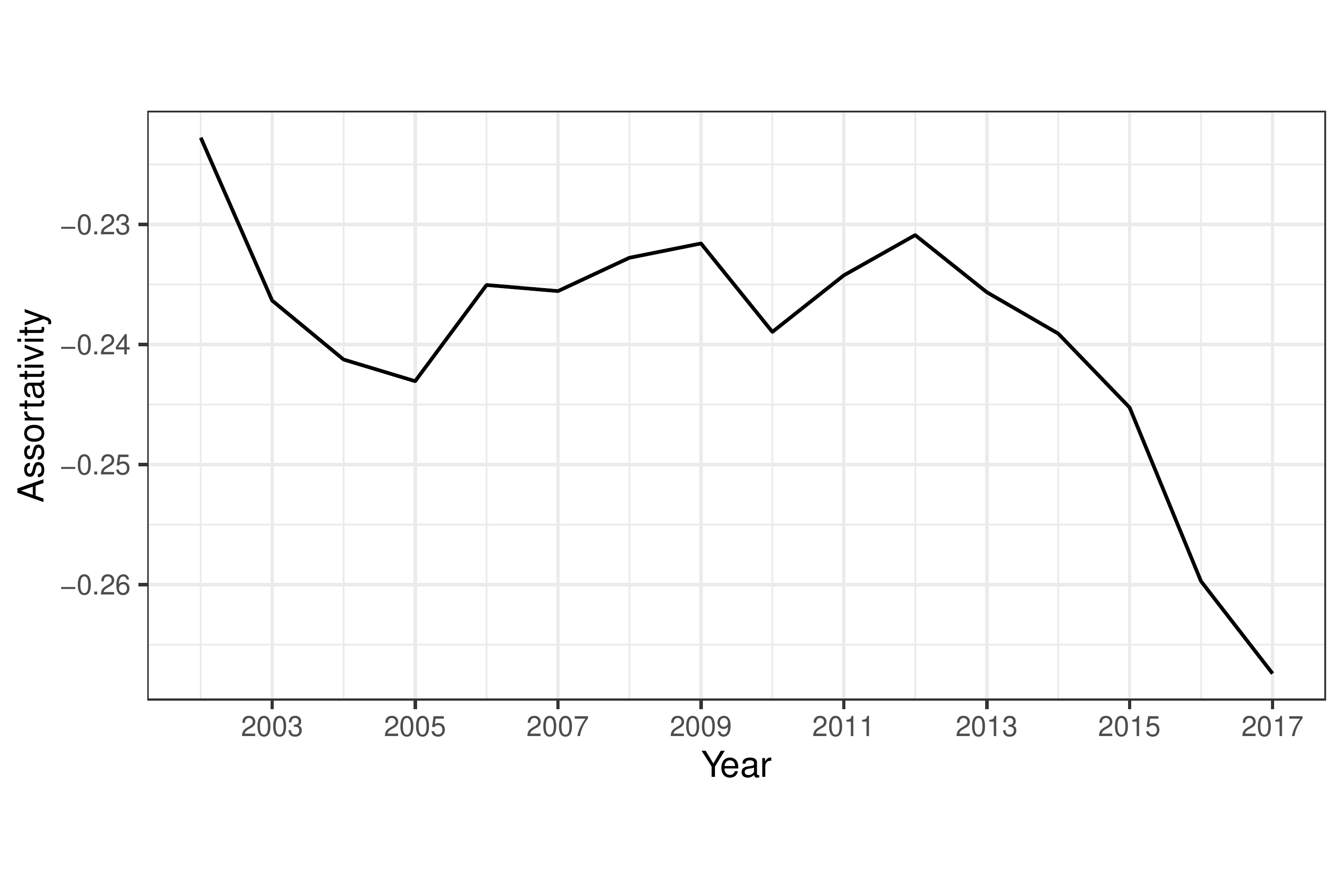}
      \caption{Assortativity}
      \label{fig:assort}
    \end{subfigure}
    \begin{subfigure}{.49\textwidth}
      \centering
      \includegraphics[width=\textwidth,clip=TRUE,trim=0cm 2cm 0cm 1.7cm]{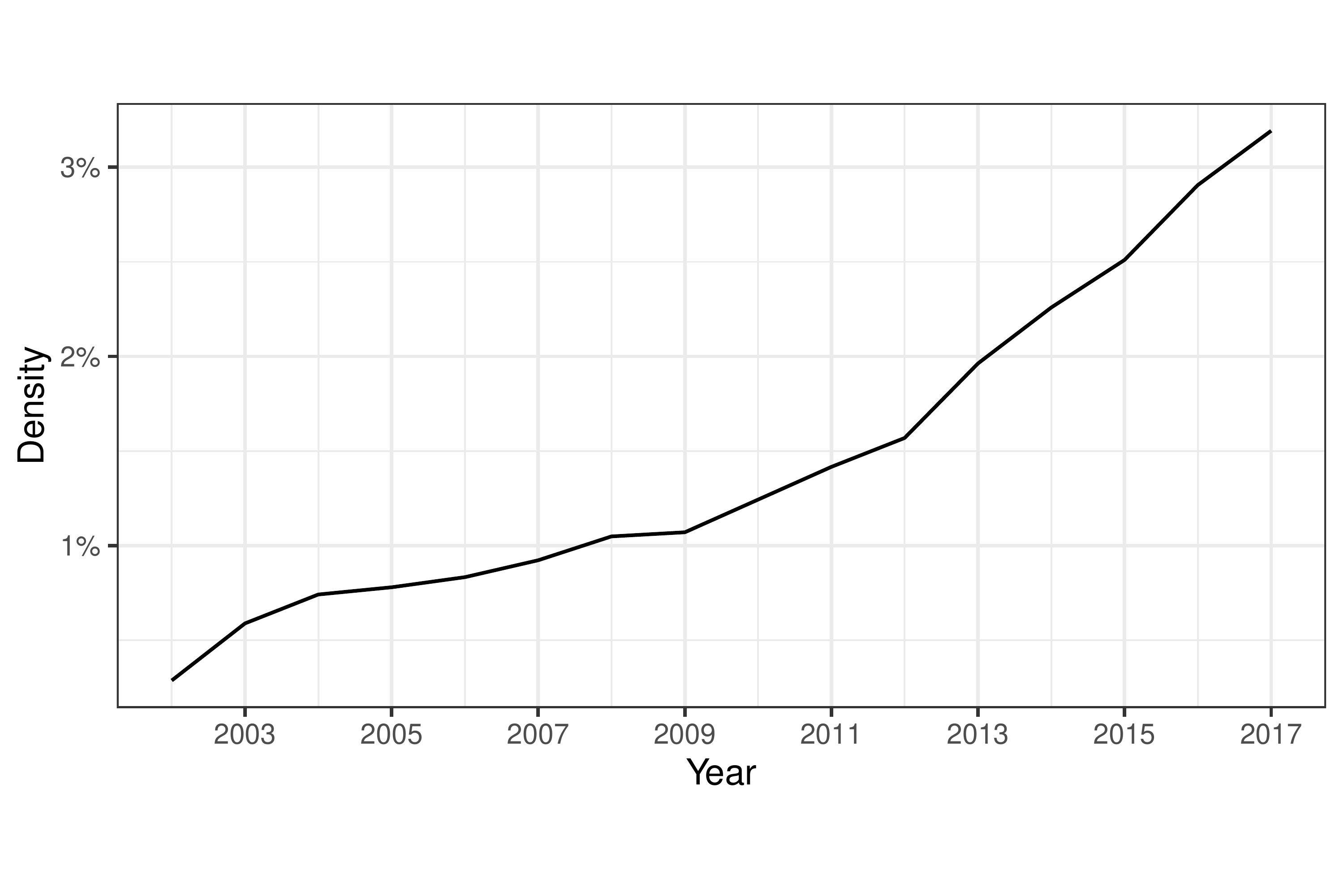}
      \caption{Density}
      \label{fig:globalNetworkMeasures-density}
    \end{subfigure}
    \caption{Network assortativity and density from the Brazilian intercity trade network between 2003 and 2017.}
\end{figure}

We start with global network measures. Figure \ref{fig:assort} portrays the network assortativity. We observe that the Brazilian network of intercity economic flows presents a disassortative mixing, with an average assortativity of $-$0.25 from 2002 to 2017. In this setting, small cities tend to connect to bigger cities, which act as hubs to the entire supply chain of Brazil. This network structure also has small network diameter, suggesting the existence of the small-world phenomenon in the network structure. That is, regardless of the geographical distances, cities tend to reside near each other in the economic sense (connections in the network).

Looking at Figure \ref{fig:assort}, we observe three different behaviors for the network assortativity. From 2002 to 2005, the network tends to evolve to a more  disassortative structure, suggesting that city hubs tend to concentrate more connections and hence become more important to the economic flows among cities in Brazil. From 2006 to 2012, we have the opposite feature: the network tends to become more assortative, meaning that connections become more dispersed throughout other cities apart from the hubs. This reduces the relative network importance of the hubs and fosters the existence of more local economic dependence rather than national dependencies warranted by the hubs, which link cities far apart from each other. Finally, from 2013 to 2017, we see a strong decrease of the network assortativity, raising the importance of city hubs to the entire national supply chain. City hubs are the most central cities in the network, which we identify in this sector later on.

Figure \ref{fig:globalNetworkMeasures-density} exhibits the network density---another global network measure---of the Brazilian network of intercity economic flows. We observe that the density ranges from roughly 0.25\% to 3.25\%. We can interpret these numbers as probabilities: if we randomly take two cities in Brazil, there is a probability that they transact with each other of 0.25\% in 2002 and 3.25\% in 2017, conditioned on the observable wire transfers. Following the rule-of-thumb in the complex network literature, this network is considered as very sparse, which corroborates the existence of city centers that link regional cities far apart. These regional cities have small degree (number of other city counterparts) and strength (intensity of transfers to other counterparts) network measure, while those centers have high degree and strength.

Table \ref{tab:centrality2003-2014} reports the ranking of the top 30 most central cities in the Brazilian network of intercity economic flows in 2003 and 2014. The network centrality is a mixed network measure because it uses not only topological network information in the direct neighborhood but also in indirect neighborhoods. We take Google PageRank as our baseline centrality measure.
The top 10 most central cities remain the same both in 2003 and 2014, suggesting that the core and most central cities persist over time. In contrast, we see more fluctuations of the other cities in the rank.

\setlength{\tabcolsep}{7pt}
\renewcommand{\arraystretch}{1.3}

\begin{sidewaystable}[htbp]
  \scriptsize
  \caption{ {\bf Ranking of 30 Municipalities with the largest centrality.}
  We considered the years 2003 (see left panel) and 2014 (see right panel). Centrality values are normalized. Therefore, we interpret then as relative values to the most central city (São Paulo in both years).}
    \begin{tabular}{p{0.7cm}p{1cm}p{2.1cm}p{2cm}p{1.2cm}||p{0.7cm}p{1cm}p{2.1cm}p{2cm}p{1.2cm}}
    \toprule\midrule
    \multicolumn{5}{c}{\textbf{2003}}     &  \multicolumn{5}{c}{\textbf{2014}} \\
    \midrule
    \textbf{Ranking} & \textbf{Centrality} & \textbf{Municipality} & \textbf{State} & \textbf{Region} & \textbf{Ranking} & \textbf{Centrality} & \textbf{Municipality} & \textbf{State} & \textbf{Region} \\
    \textbf{1} & 100\% & São Paulo & São Paulo & Southeast &        \textbf{1} & 100\% & São Paulo & São Paulo & Southeast \\
    \textbf{2} & 73\%  & Rio de Janeiro & Rio de Janeiro & Southeast        & \textbf{2} & 69\%  & Rio de Janeiro & Rio de Janeiro & Southeast \\
    \textbf{3} & 15\%  & Barueri & São Paulo & Southeast        & \textbf{3} & 25\%  & Brasília & Distrito Federal & Midwest \\
    \textbf{4} & 14\%  & Brasília & Distrito Federal & Midwest &        \textbf{4} & 14\%  & Osasco & São Paulo & Southeast \\
    \textbf{5} & 12\%  & Belo Horizonte & Minas Gerais & Southeast       & \textbf{5} & 11\%  & Barueri & São Paulo & Southeast \\
    \textbf{6} & 8\%   & Porto Alegre & Rio Grande do Sul & South &        \textbf{6} & 9\%   & Belo Horizonte & Minas Gerais & Southeast \\
    \textbf{7} & 7\%   & Curitiba & Paraná & South &       \textbf{7} & 8\%   & Curitiba & Paraná & South \\
    \textbf{8} & 7\%   & Campinas & São Paulo & Southeast &        \textbf{8} & 6\%   & Porto Alegre & Rio Grande do Sul & South \\
    \textbf{9} & 6\%   & Osasco & São Paulo & Southeast &        \textbf{9} & 5\%   & Salvador & Bahia & Northeast \\
    \textbf{10} & 6\%   & Belém & Pará  & North &        \textbf{10} & 4\%   & São Caetano do Sul & São Paulo & Southeast \\
    \textbf{11} & 5\%   & São Bernardo do Campo & São Paulo & Southeast        & \textbf{11} & 4\%   & Recife & Pernambuco & Northeast \\
    \textbf{12} & 4\%   & São Caetano do Sul & São Paulo & Southeast        & \textbf{12} & 4\%   & Fortaleza & Ceará & Northeast \\
    \textbf{13} & 4\%   & Jaguariúna & São Paulo & Southeast &        \textbf{13} & 3\%   & Campinas & São Paulo & Southeast \\
    \textbf{14} & 4\%   & Salvador & Bahia & Northeast &        \textbf{14} & 3\%   & São Bernardo do Campo & São Paulo & Southeast \\
    \textbf{15} & 3\%   & Recife & Pernambuco & Northeast &        \textbf{15} & 3\%   & Manaus & Amazonas & North \\
    \textbf{16} & 3\%   & Manaus & Amazonas & North &        \textbf{16} & 3\%   & Goiânia & Goiás & Midwest \\
    \textbf{17} & 3\%   & Betim & Minas Gerais & Southeast &        \textbf{17} & 2\%   & Vitória & Espírito Santo & Southeast \\
    \textbf{18} & 3\%   & Vitória & Espírito Santo & Southeast &        \textbf{18} & 2\%   & Guarulhos & São Paulo & Southeast \\
    \textbf{19} & 2\%   & Goiânia & Goiás & Midwest &        \textbf{19} & 2\%   & Camaçari & Bahia & Northeast \\
    \textbf{20} & 2\%   & Fortaleza & Ceará & Northeast &        \textbf{20} & 2\%   & Itajaí & Santa Catarina & South \\
    \textbf{21} & 2\%   & Niterói & Rio de Janeiro & Southeast &        \textbf{21} & 2\%   & Florianópolis & Santa Catarina & South \\
    \textbf{22} & 2\%   & Florianópolis & Santa Catarina & South &        \textbf{22} & 2\%   & Gaspar & Santa Catarina & South \\
    \textbf{23} & 2\%   & Santo André & São Paulo & Southeast &        \textbf{23} & 2\%   & São Luís & Maranhão & Northeast \\
    \textbf{24} & 2\%   & Camaçari & Bahia & Northeast &        \textbf{24} & 2\%   & Betim & Minas Gerais & Southeast \\
    \textbf{25} & 2\%   & Guarulhos & São Paulo & Southeast &        \textbf{25} & 2\%   & Duque de Caxias & Rio de Janeiro & Southeast \\
    \textbf{26} & 2\%   & São José dos Campos & São Paulo & Southeast &        \textbf{26} & 2\%   & Santo André & São Paulo & Southeast \\
    \textbf{27} & 2\%   & Uberlândia & Minas Gerais & Southeast &        \textbf{27} & 2\%   & Cuiabá & Mato Grosso & Midwest \\
    \textbf{28} & 2\%   & João Pessoa & Paraíba & Northeast &        \textbf{28} & 1\%   & Contagem & Minas Gerais & Southeast \\
    \textbf{29} & 1\%   & Poá   & São Paulo & Southeast &        \textbf{29} & 1\%   & Teresina & Piauí & Northeast \\
    \textbf{30} & 1\%   & Natal & Rio Grande do Norte & Northeast        & \textbf{30} & 1\%   & Uberlândia & Minas Gerais & Southeast \\
    \midrule\bottomrule
    \end{tabular}%
  \label{tab:centrality2003-2014}%
\end{sidewaystable}%

\begin{figure}
    \centering
    \begin{subfigure}{.49\textwidth}
      \centering
      \includegraphics[width=\textwidth,clip=TRUE,trim=0.2cm 0.8cm 0cm 0.3cm]{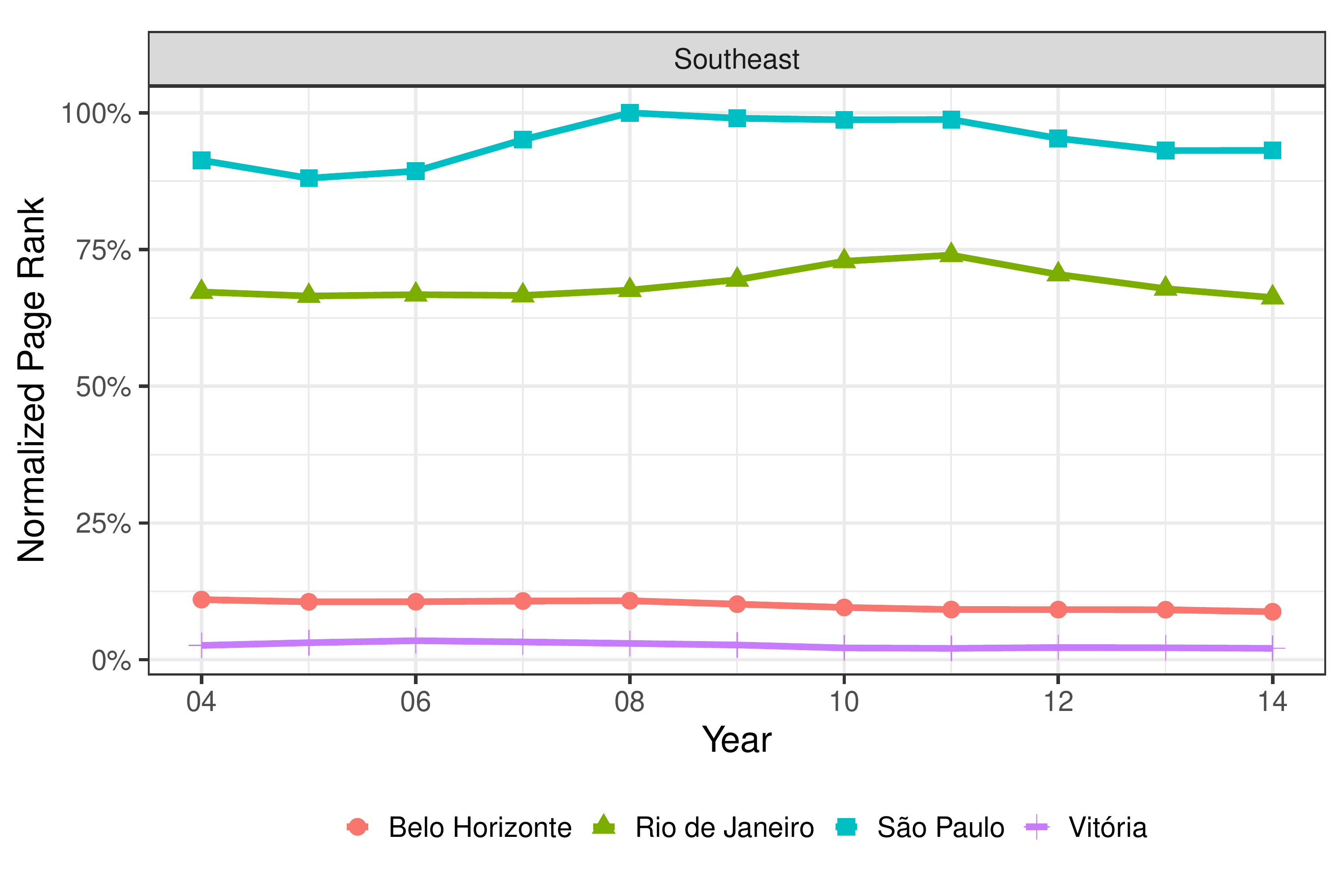}
      \caption{Southeast}
      \label{fig:nc-Sudeste}
    \end{subfigure}
    \begin{subfigure}{.49\textwidth}
      \centering
      \includegraphics[width=\textwidth,clip=TRUE,trim=0.2cm 0.8cm 0cm 0.3cm]{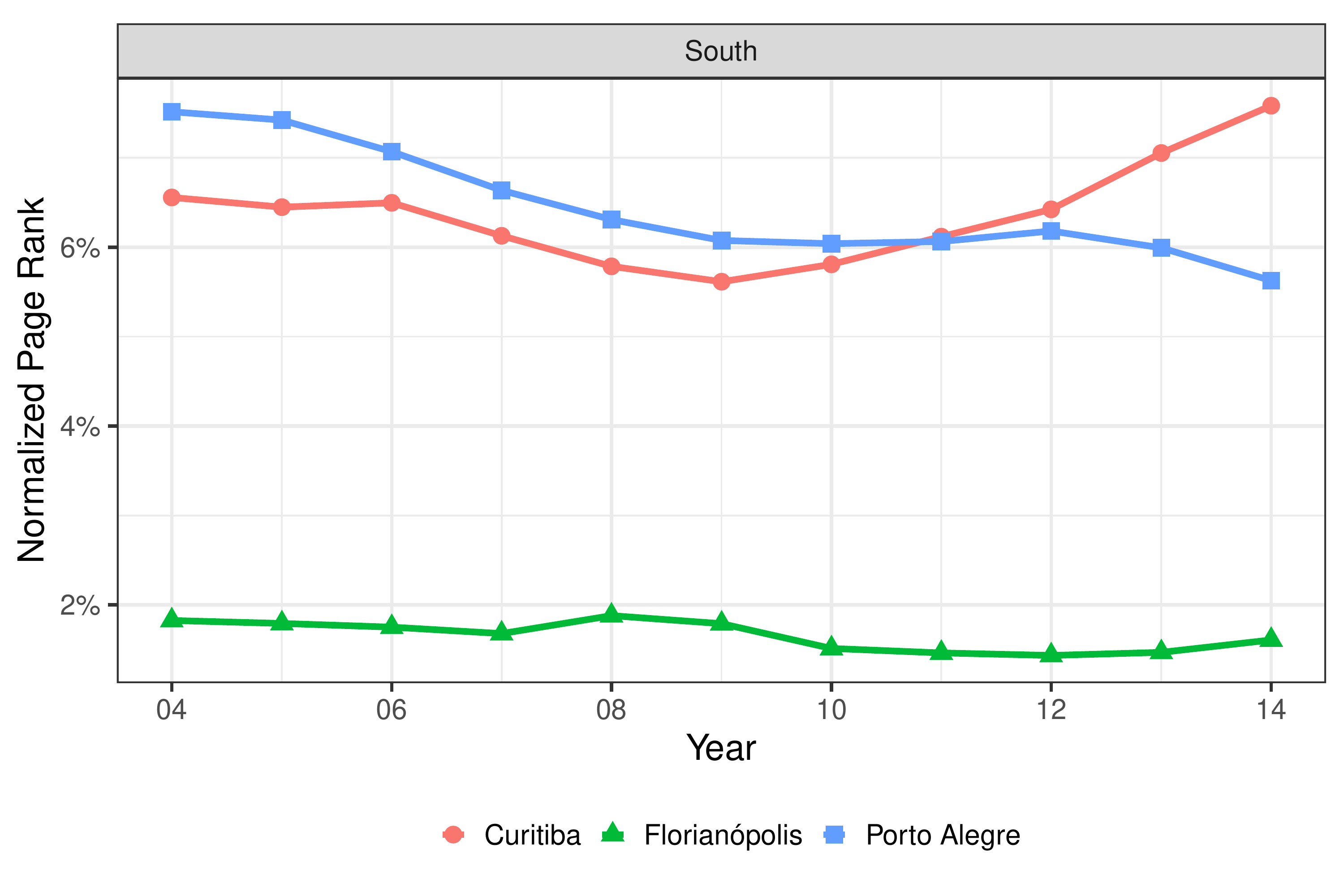}
      \caption{Midwest}
      \label{fig:nc-Sul}
    \end{subfigure}

    \begin{subfigure}{.49\textwidth}
      \centering
      \includegraphics[width=\textwidth,clip=TRUE,trim=0.2cm 0.8cm 0cm 0.3cm]{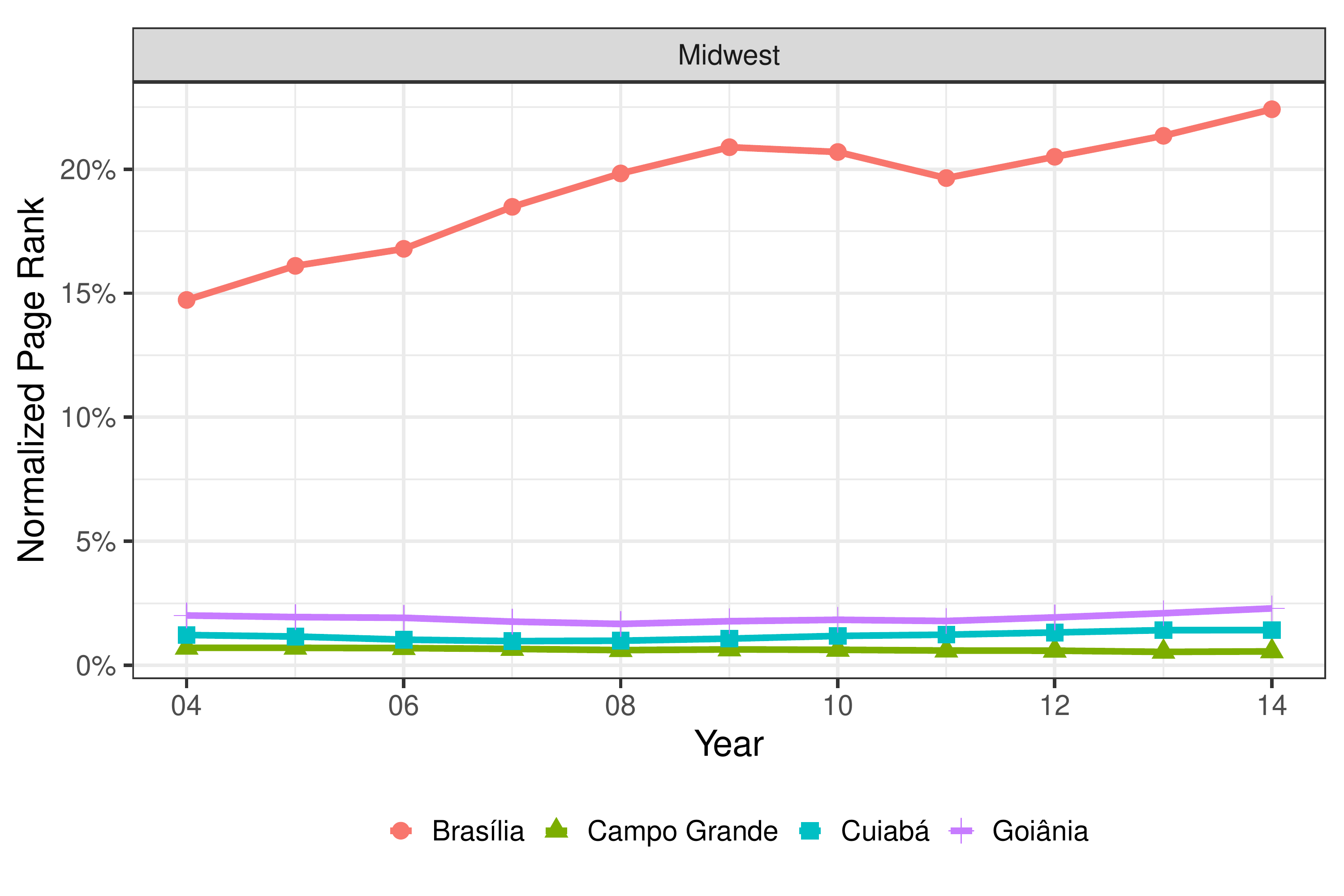}
      \caption{Midwest}
      \label{fig:nc-Centroeste}
    \end{subfigure}
    \begin{subfigure}{.49\textwidth}
      \centering
      \includegraphics[width=\textwidth,clip=TRUE,trim=0.2cm 0.8cm 0cm 0.3cm]{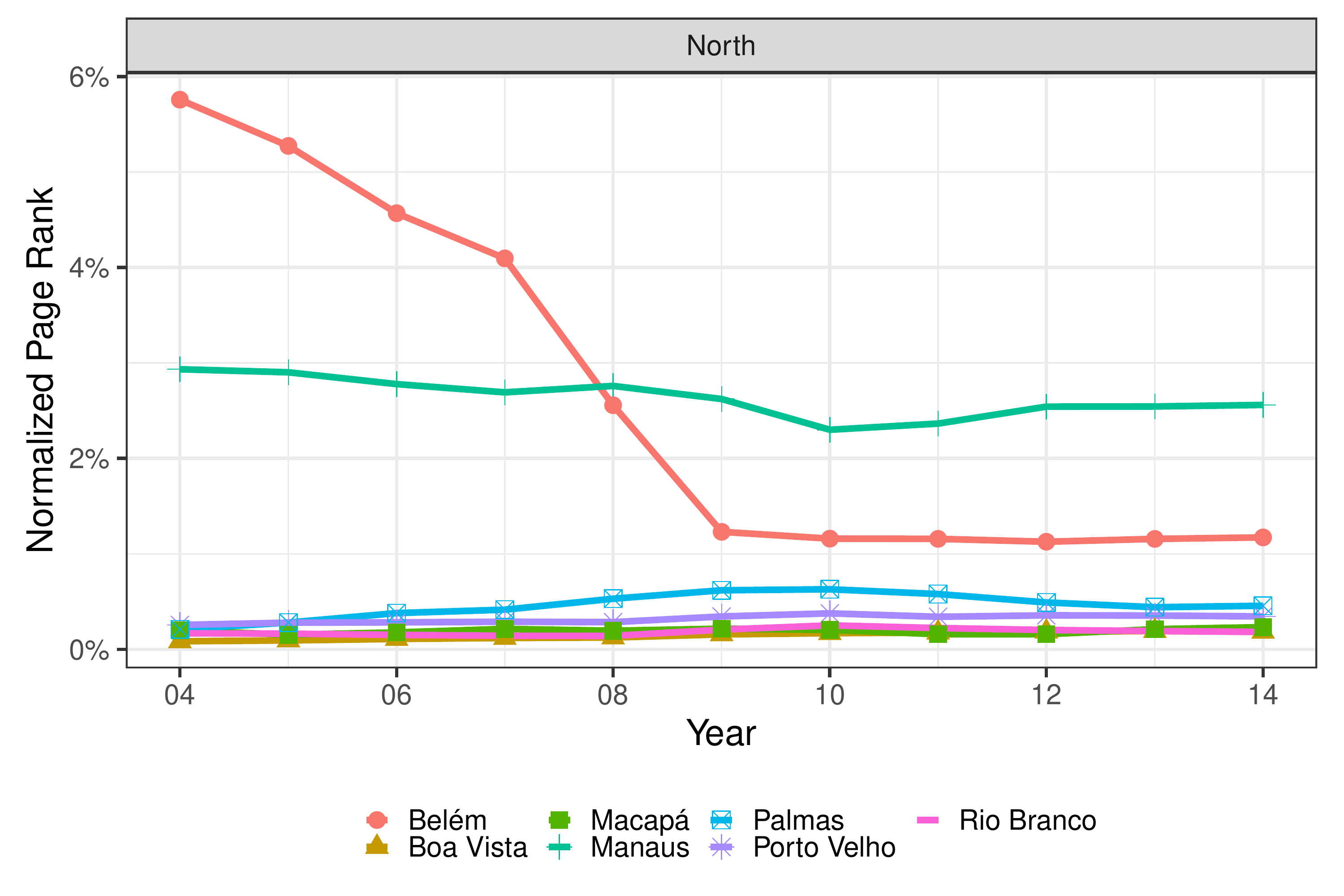}
      \caption{North}
      \label{fig:nc-Norte}
    \end{subfigure}

    \begin{subfigure}{\textwidth}
      \centering
      \includegraphics[width=0.7\textwidth,clip=TRUE,trim=0.2cm 0.8cm 0cm 0.3cm]{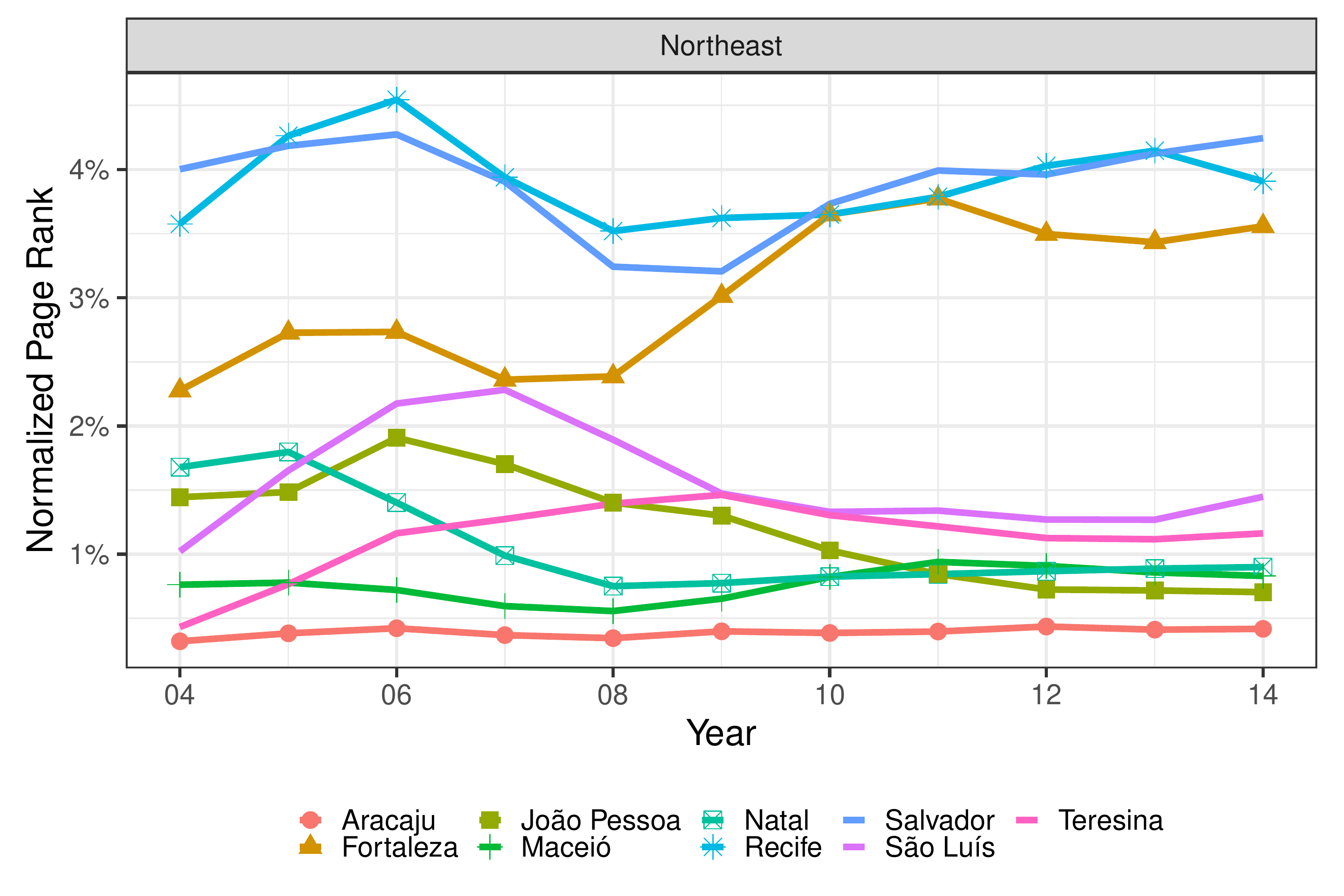}
      \caption{Northeast}
      \label{fig:nc-Nordeste}
    \end{subfigure}
    \caption{Evolution of network centrality for Brazilian capitals of the (a) Southeast, (b) South, and (c) Midwest regions. We evaluate network centrality using the downstream perspective, i.e., links follow the money flow. Data is smoothed with a time window of two years from 2003--2014.}
\end{figure}

City capitals are the ones that tend to connect cities far apart and therefore are the most probable candidates of being hubs of the supply chain network in Brazil. Figures \ref{fig:nc-Sudeste}--\ref{fig:nc-Nordeste} show the evolution of the network centrality for the capitals of the Southeast, South, Midwest, North, and Northeast, respectively. The South and Southeast are the most developed  in Brazil, while the Midwest contains Brasília, which is the federal district of Brazil. The North and Northeast are the least developed regions in Brazil. São Paulo is by far the most central city in Brazil, followed by Rio de Janeiro, both located in the Southeast region. Their relative importance to the entire network remains stable over time, evidencing their importance to the entire supply chain in Brazil.


The relative importance of some capitals change over time. For instance, in the South region, the relative importance of Curitiba quickly increases after 2011, while that of Porto Alegre decreases in the same period. In the North, Belém remained as the most important regional center until 2008, after which Manaus took the top 1 place in the region. All the other capitals in the North have very small importance, suggesting that the link of the North region to the remainder of the country goes through Manaus and Belém.  In the Northeast region, we observe a large heterogeneity. While we observe a steady and large importance for Recife and Salvador, the importance of Fortaleza quickly increases after 2007. The role of São Luis and João Pessoa in the network decreases before 2008, and remains rather stable after then. In the Midwest, we observe a quick increase of the importance of Brasília throughout the entire sample period.

\begin{figure}
    \begin{center}
      \includegraphics[width=\textwidth,clip=TRUE,trim=0.2cm 1cm 0.2cm 0.8cm]{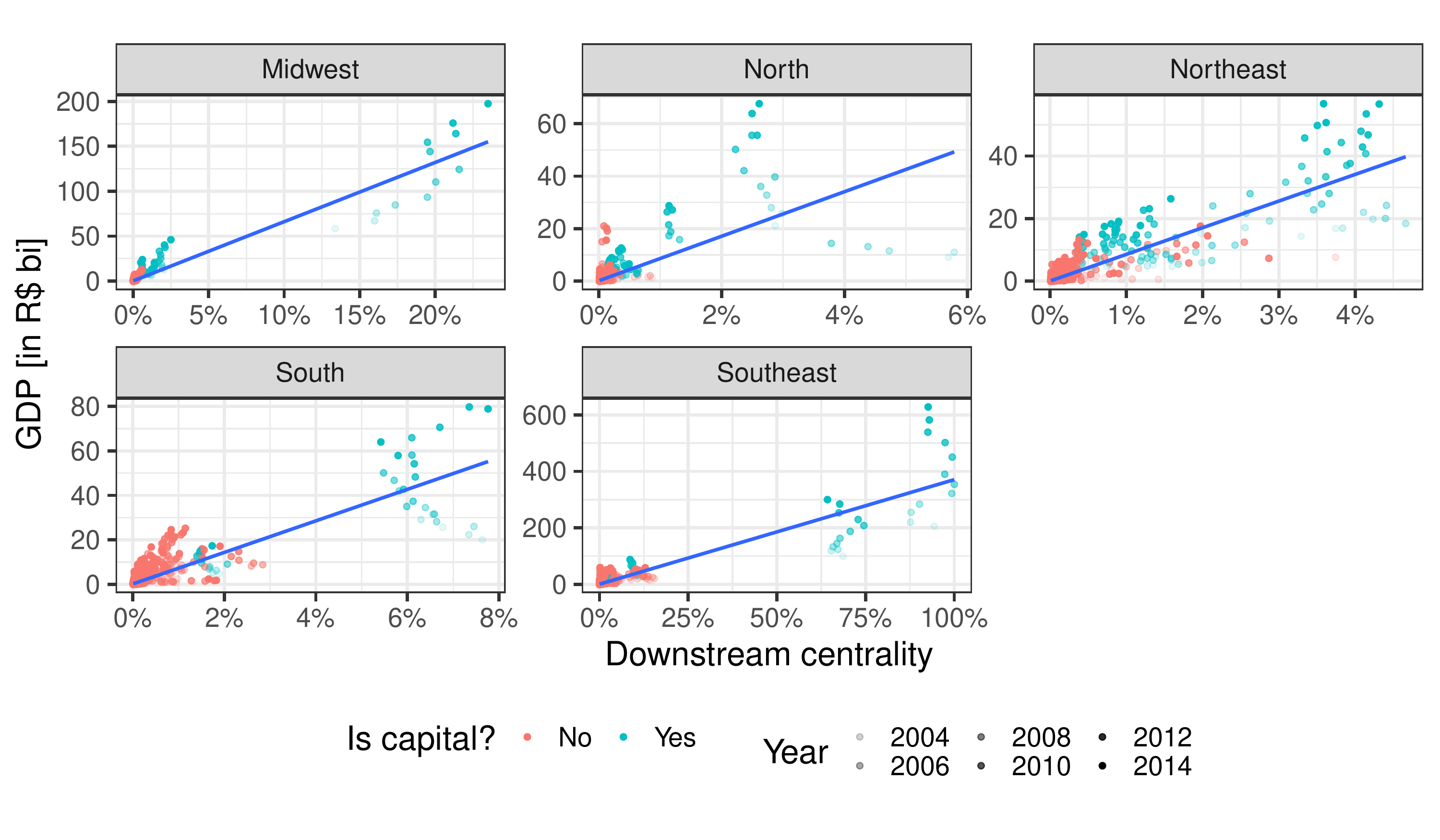}
           \caption{Scatterplot that compares downstream centrality and size (measured by city GDP) for each year. We also include information on whether the city is a capital of a State in Brazil. }
           \label{fig:scat}
    \end{center}
\end{figure}

Figure \ref{fig:scat} presents a scatterplot of the downstream centrality versus city size (measured using GDP), for specific years (the years are reflected by the shading of the observation, with more recent years being darker). If the city is a capital, then the color is blue and, if it is not, the color is red. By fitting a standard OLS curve, we find a positive correlation between city size and downstream centrality, suggesting that larger cities are more central. Such relationship tends to increase over time and is invariant over different Brazilian regions. In particular, capitals seem to affect this relationship by making it stronger.

\begin{figure}
    \begin{center}
      \includegraphics[width=0.8\textwidth,clip=TRUE,trim=2cm 0.3cm 0cm 0.4cm]{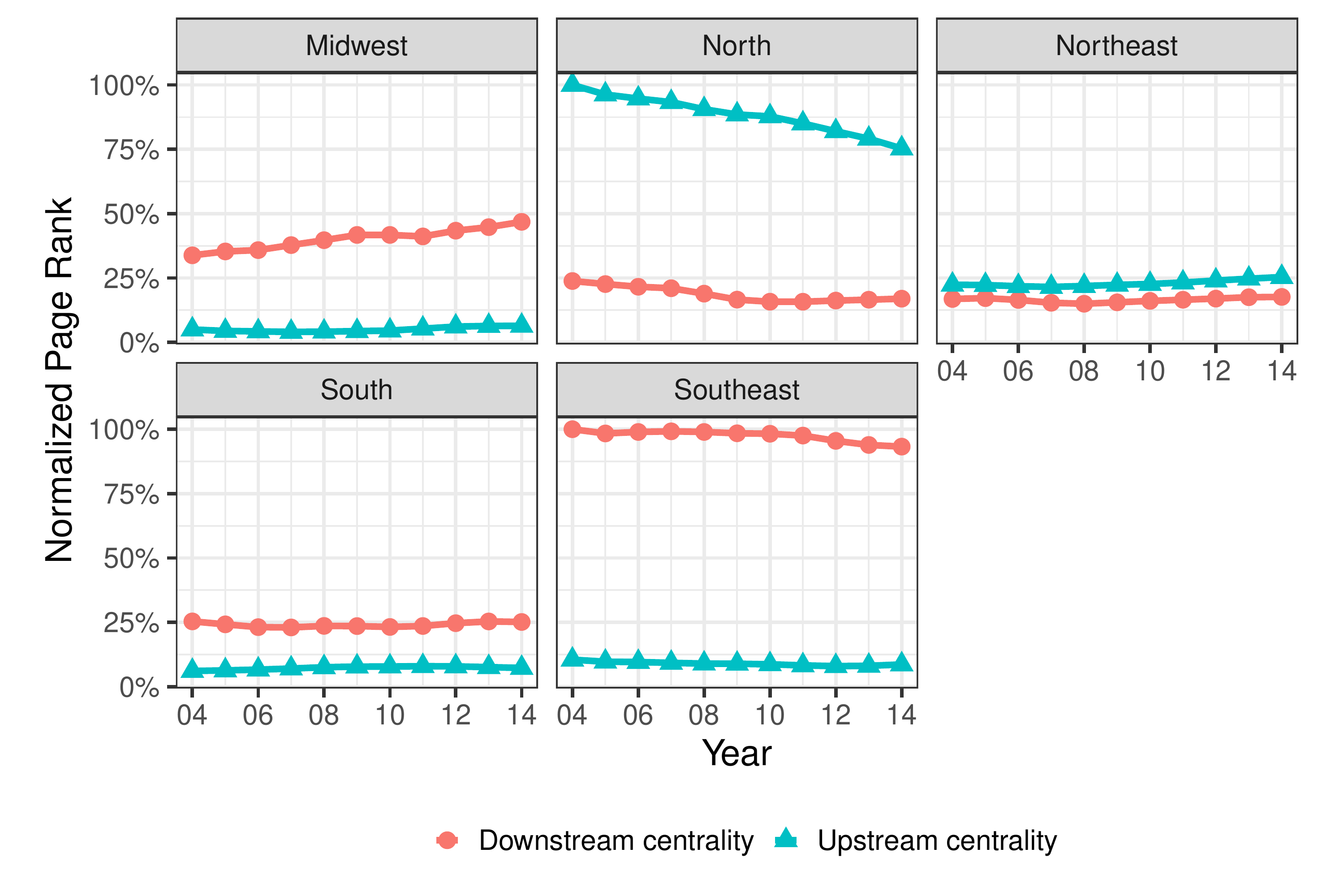}
       \caption{Evolution over time of normalized PageRank - Upstream and Downstream Centrality for each Region in Brazil. The red and blue lines stand for downstream and upstream centrality, respectively}
       \label{fig:up_down}
    \end{center}
\end{figure}

We present the evolution over time of the average upstream and downstream centralities for each Brazilian region in Figure \ref{fig:up_down}. Our results highlight major differences across Brazil. Downstream Centrality is high over this time period for the Southeast, whereas upstream centrality is high for the Northern Region (although there is a downward trend over time). There is also an upward trend in the downstream centrality in the Midwest (almost doubles in a decade).

Table \ref{tab:DOEC-DOES-2014} shows the top 30 cities with largest dependence on external customers and suppliers in 2014, respectively. These measures are strictly local. That is, their economic relies on customers buying or suppliers selling products/services in other cities rather than within the same city. A large external dependence corroborates the Ricardian theory of comparative advantage in which cities should specialize in the production of those goods and services in which they enjoy comparative advantage. In this case, we would observe a large economic integration among cities. Overall, we observe a high external dependence, allowing us to conclude for the existence of a strong economic integration across Brazilian cities.

\setlength{\tabcolsep}{7pt}
\renewcommand{\arraystretch}{1.3}

\begin{sidewaystable}[htbp]
  \centering
  \scriptsize
  \caption{{\bf Ranking of the 30 Municipalities with the largest dependence on external customers and on suppliers}. The dependence external customers (DOEC) is shown in the left panel and the dependence  on suppliers (DOES) is shown in the right panel. This table shows the data obtained for 2014.}
    \begin{tabular}{p{0.7cm}p{1cm}p{3cm}p{2cm}p{1.2cm}||p{0.7cm}p{1cm}p{3cm}p{2cm}p{1.2cm}}
    \toprule\midrule
    \multicolumn{5}{c}{\textbf{Dependence on External Customers (DOEC) in 2014}} & \multicolumn{5}{c}{\textbf{Dependence on External Suppliers (DOES) in 2014}} \\
    \midrule
    \textbf{Ranking} & \textbf{DOEC} & \textbf{Municipality} & \textbf{State} & \textbf{Region} &        \textbf{Ranking} & \textbf{DOES} & \textbf{Municipality} & \textbf{State} & \textbf{Region} \\
    1     & 100.00\% & Cachoeira Grande & Maranhão & Northeast &        1     & 100.00\% & Cachoeira Grande & Maranhão & Northeast \\
    2     & 100.00\% & São Gonçalo do Gurguéia & Piauí & Northeast        & 2     & 100.00\% & São Gonçalo do Gurguéia & Piauí & Northeast \\
    3     & 100.00\% & Jardim de Angicos & Rio Grande do Norte & Northeast &        3     & 100.00\% & Jardim de Angicos & Rio Grande do Norte & Northeast \\
    4     & 100.00\% & Vila Flor & Rio Grande do Norte & Northeast &        4     & 100.00\% & Vila Flor & Rio Grande do Norte & Northeast \\
    5     & 100.00\% & Curral Velho & Paraíba & Northeast &        5     & 100.00\% & Curral Velho & Paraíba & Northeast \\
    6     & 100.00\% & São José de Princesa & Paraíba & Northeast &        6     & 100.00\% & São José de Princesa & Paraíba & Northeast \\
    7     & 100.00\% & Serra Redonda & Paraíba & Northeast &        7     & 100.00\% & Serra Redonda & Paraíba & Northeast \\
    8     & 99.71\% & Tupirama & Tocantins & North        & 8     & 99.79\% & Pau D'Arco do Piauí & Piauí & Northeast \\
    9     & 99.62\% & Serra do Navio & Amapá & North &        9     & 99.49\% & São Domingos do Cariri & Paraíba & Northeast \\
    10    & 99.55\% & Pau D'Arco do Piauí & Piauí & Northeast       & 10    & 98.63\% & São Roberto & Maranhão & Northeast \\
    11    & 99.33\% & Peixe & Tocantins & North        & 11    & 98.62\% & Tupirama & Tocantins & North \\
    12    & 99.28\% & Parari & Paraíba & Northeast        & 12    & 98.46\% & Marajá do Sena & Maranhão & Northeast \\
    13    & 99.00\% & Piau  & Minas Gerais & Southeast        & 13    & 98.27\% & Pedro Laurentino & Piauí & Northeast \\
    14    & 98.65\% & Passagem Franca do Piauí & Piauí & Northeast        & 14    & 98.23\% & Matões do Norte & Maranhão & Northeast \\
    15    & 98.65\% & São Miguel das Matas & Bahia & Northeast        & 15    & 98.19\% & Lagoa d'Anta & Rio Grande do Norte & Northeast \\
    16    & 98.40\% & Pedro Laurentino & Piauí & Northeast       & 16    & 97.81\% & Matinhas & Paraíba & Northeast \\
    17    & 98.35\% & Treviso & Santa Catarina & South &       17    & 97.79\% & Camacho & Minas Gerais & Southeast \\
    18    & 98.18\% & Rafard & São Paulo & Southeast &       18    & 97.71\% & Santo Antônio do Retiro & Minas Gerais & Southeast \\
    19    & 98.09\% & Água Nova & Rio Grande do Norte & Northeast       & 19    & 97.69\% & Água Nova & Rio Grande do Norte & Northeast \\
    20    & 98.02\% & Salmourão & São Paulo & Southeast &    20    & 97.42\% & Itaquara & Bahia & Northeast \\
    21    & 97.97\% & Salgado de São Félix & Paraíba & Northeast &        21    & 97.35\% & Gentio do Ouro & Bahia & Northeast \\
    22    & 97.65\% & Major Gercino & Santa Catarina & South &        22    & 97.32\% & Parari & Paraíba & Northeast \\
    23    & 97.61\% & Quatá & São Paulo & Southeast &        23    & 97.23\% & Bernardino Batista & Paraíba & Northeast \\
    24    & 97.51\% & Viçosa & Rio Grande do Norte & Northeast        & 24    & 97.13\% & Gado Bravo & Paraíba & Northeast \\
    25    & 97.43\% & Mutuípe & Bahia & Northeast &        25    & 97.12\% & Lagoinha do Piauí & Piauí & Northeast \\
    26    & 97.41\% & Jenipapo de Minas & Minas Gerais & Southeast &        26    & 96.73\% & Rafael Godeiro & Rio Grande do Norte & Northeast \\
    27    & 97.25\% & São Francisco do Pará & Pará  & North &        27    & 96.59\% & Major Sales & Rio Grande do Norte & Northeast \\
    28    & 97.15\% & Bom Jardim da Serra & Santa Catarina & South &        28    & 96.37\% & Loreto & Maranhão & Northeast \\
    29    & 97.14\% & São Miguel da Baixa Grande & Piauí & Northeast &        29    & 96.36\% & Santo André & Paraíba & Northeast \\
    30    & 97.10\% & Areia de Baraúnas & Paraíba & Northeast        & 30    & 96.24\% & Santana da Ponte Pensa & São Paulo & Southeast \\
    \midrule\bottomrule
    \end{tabular}%
  \label{tab:DOEC-DOES-2014}%
\end{sidewaystable}%

\begin{figure}
    \begin{center}
      \includegraphics[width=\textwidth,clip=TRUE,trim=0cm 0.8cm 0cm 0.3cm]{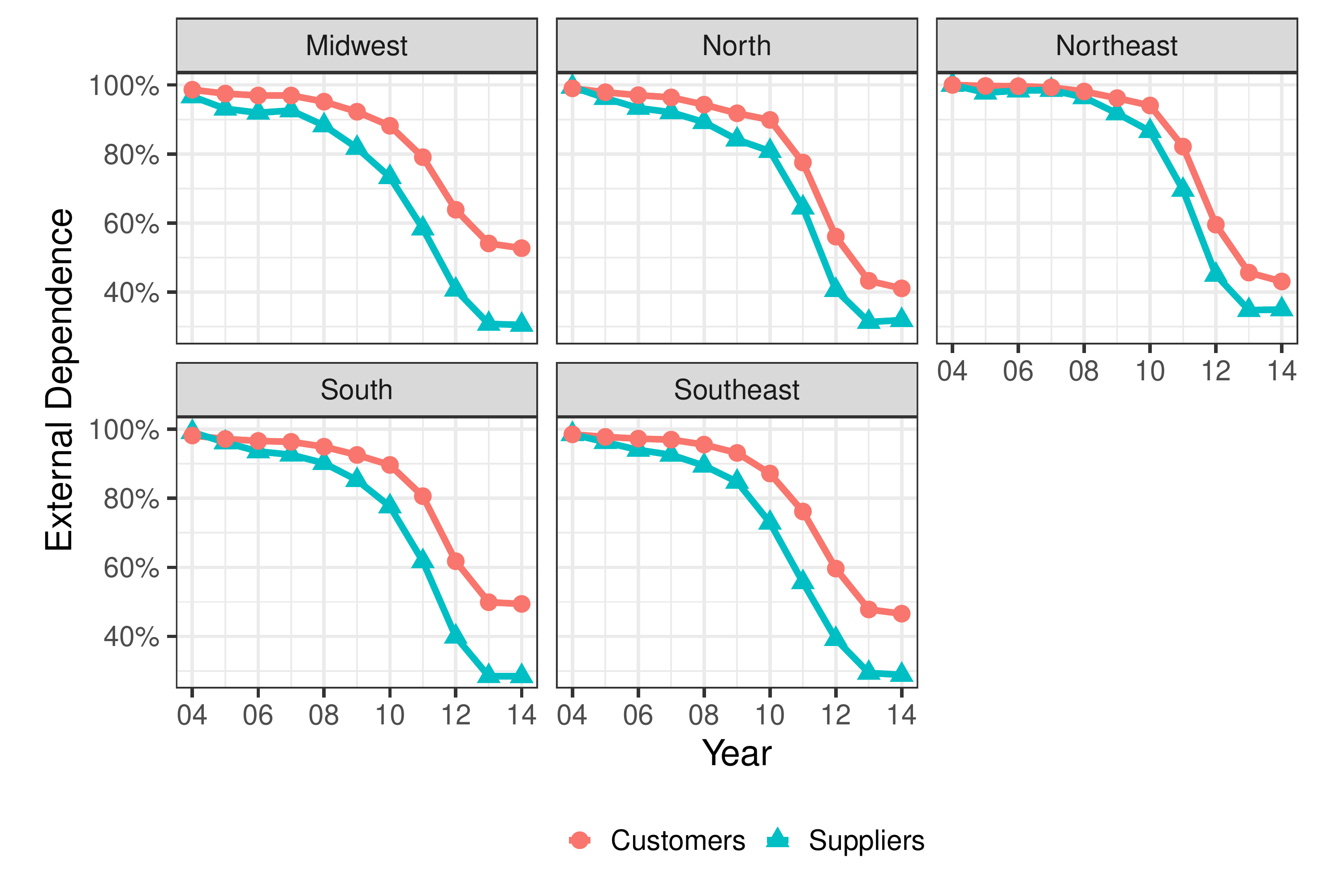}
       \caption{Evolution of the average regional dependence on external customers (red or circle-shaped curve) and suppliers (blue or triangle-shaped curve) for cities for the period 2003--2014.}
         \label{fig:ext}
         \end{center}
\end{figure}

Figure \ref{fig:ext} presents the evolution of dependence on external customers and suppliers. In the beginning of our sample, the dependence on external customers and suppliers in different regions of Brazil was high. Over time, this dependence substantially reduces. The dependence on external customers is still high in some regions, such as in the Midwest, South and Southeast (approximately 50\%). An additional finding is that average regional dependence on external suppliers is lower than that on customers, suggesting that there was strong investments in reducing the dependence on external suppliers in Brazilian regions.

\begin{figure}
    \centering
    \begin{subfigure}{\textwidth}
      \centering
      \includegraphics[width=\textwidth,clip=TRUE,trim=0.2cm 1.2cm 0.2cm 0.2cm]{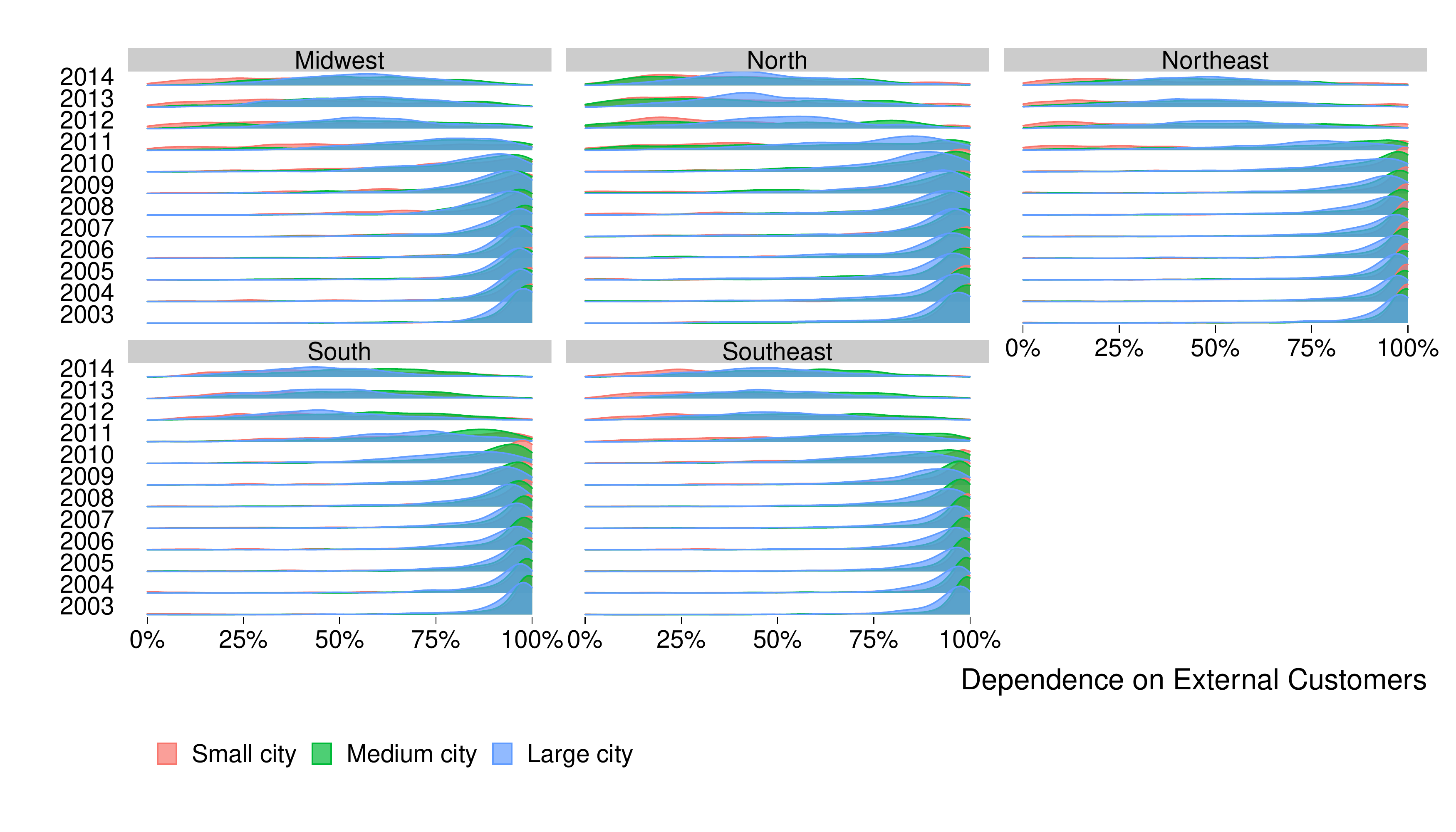}
      \caption{External Customers}
      \label{fig:cust}
    \end{subfigure}

    \begin{subfigure}{\textwidth}
      \centering
      \includegraphics[width=\textwidth,clip=TRUE,trim=0.2cm 1.2cm 0.2cm 0.2cm]{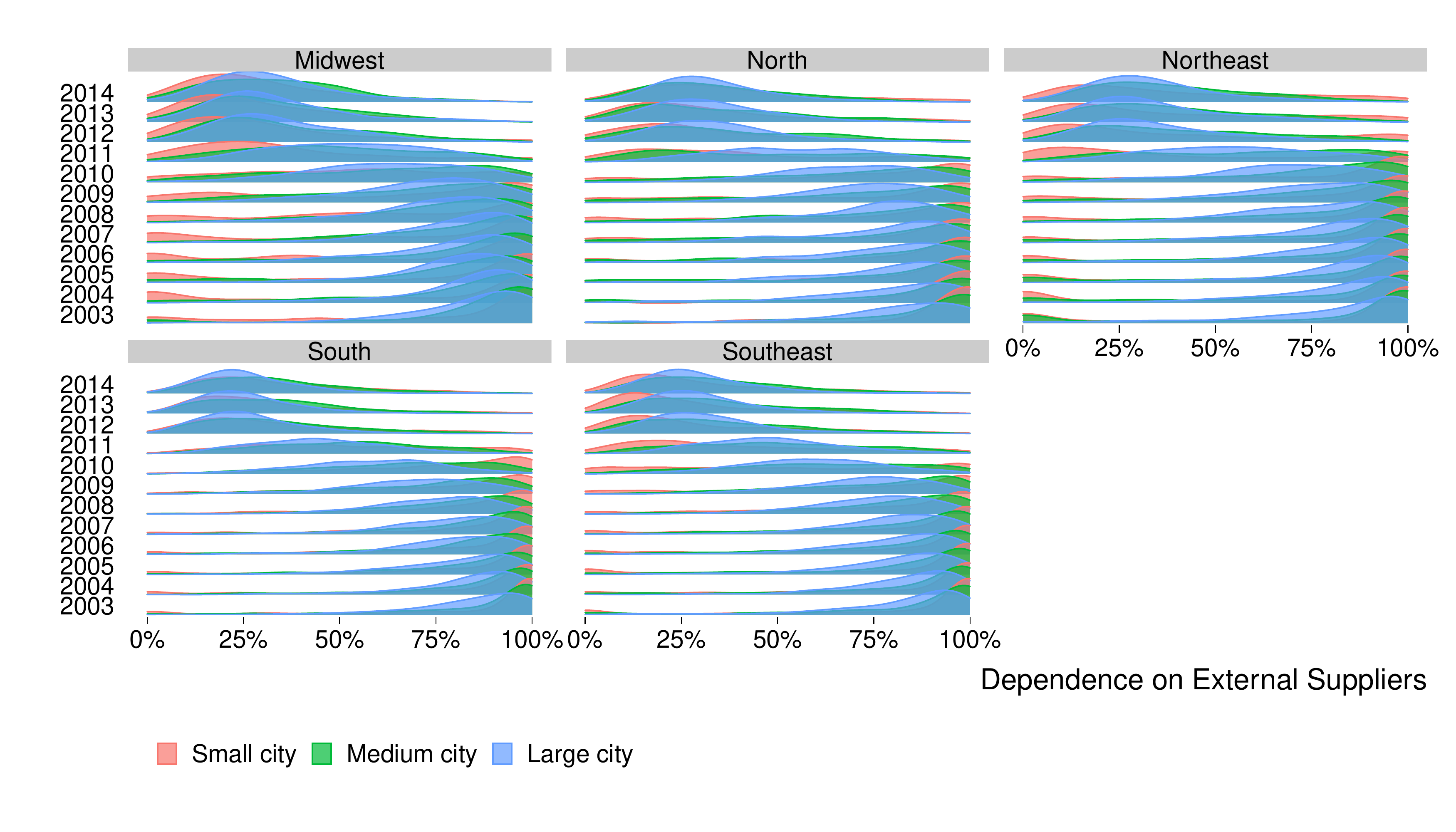}
      \caption{External Suppliers}
      \label{fig:sup}
    \end{subfigure}
    \caption{Evolution of the city-specific distribution of dependence on external (a) customers and (b) suppliers in the period 2003--2014.}
\end{figure}

We can also analyze how the entire city-specific distribution of dependence on external customers and supply behave over time in Figures \ref{fig:cust} and \ref{fig:sup}, respectively. We segregate by city size in accordance with the city population distribution.\footnote{Small, medium, and large cities correspond to the bottom, middle, and upper terciles of the regional-specific city GDP distribution.} The dependence on external customers reduces over the sample period. This trend is robust across regions in Brazil. However, when we compare small, medium and large cities, we find that the reduction in this dependence is stronger in small cities for all regions. This is also true for medium cities in the North region. However, in other regions such as the South and Southeast, medium cities are still highly dependent on external customers.

Looking now at the dependence on external suppliers, there is also a trend suggesting the reduction of external dependence of large cities, although the results are more heterogeneous in this case. In the Midwest, Southeast and Northeast, small cities became even less dependent on external suppliers, which suggests that they are more diversified in the present than they were before. There is also a reduction in the external dependence of medium cities, but not as strong as the one occurred in small and large cities.

\section{Socioeconomic, legal, and demographic drivers of the wire transfers network}

In this section, we apply an econometric exercise to understand how socioeconomic, legal and demography characteristics of Brazilian municipalities correlate with network centrality (PageRank), degree (number of suppliers and customers), and strength (total received and paid). Such analysis permits us understand the underlying drivers of these measures. We match  our city-level wire transfers data set explored in the previous section with several socioeconomic, Judiciary and demography databases. We explore each of them as we discuss the econometric specifications.

To understand the correlates that drive city-level network measures, we use the following panel-data specification:

\begin{align}
    y_{irt} = \alpha_{rt} + \bm{\beta}^T\bm{X}_{irt} + \epsilon_{irt},
    \label{eq:spec}
\end{align}

\noindent in which $i$, $r$, and $t$ index city, region (of the city), and time (in years). We use as outcome variables $y_{irt}$ the following city-level network measures: downstream and upstream PageRank centrality, in-degree (number of customers), out-degree (number of suppliers), in-strength (total received), and out-strength (total paid). The factor $\epsilon_{irt}$ is the standard error term. We use robust error clustering at the city-level in all our specifications. The term $\alpha_{rt}$ is a region-time fixed effects, whose introduction enables comparisons of municipalities only of the same Brazilian region. Since Brazil has  continental dimensions with five regions (North, Northeast, Midwest, Southeast, and South) with very different socioeconomic, demographic, and cultural aspects, the within-region comparison mitigates concerns about omitted variables related to non-observable time-variant characteristics of these five regions.

The vector $\bm{X}$ contains all the socioeconomic, legal and demographic characteristics that could drive city outcome variables. We use a broad set of regressors, such as to have a comprehensive picture of the underlying drivers of the network measures of Brazilian cities. We use the following regressors:

\begin{itemize}

    \item \textit{Municipal GDP}: we believe that the city's size can have a strong role on some network measures. For instance, bigger cities are more likely to intermediate more economic transactions, have more firms, and perform more  transactions with firms in other cities. This data is public and comes from the \textit{Instituto Brasileiro de Geografia e Estatística} (IBGE), which is the agency responsible for official collection of statistical, geographic, cartographic, geodetic and environmental information in Brazil.

    \item \textit{Share of the total municipal exports to its GDP}: we introduce this variable to capture any differences of the impact of domestic- and foreign-oriented economies on the city-level network measures. Exporting cities may have an economic local machinery and needs that are fundamentally different from those domestic-oriented cities. Since our network captures domestic economic transactions among different cities, such economy orientation may have an impact on how and to what extent exporting cities engage in the national supply chain comparatively to domestic-oriented cities. This data is public and comes from the \textit{Ministério da Indústria, Comércio Exterior e Serviços}, which is responsible for the development policy on industry, trade and services, and foreign trade policies.

    \item \textit{Share of total municipal bank credit to its GDP}: this variable controls for the local financial development of cities. There is a large body in the finance-nexus literature that advocates that financial development correlates with economic growth. Since our network is built from economic activity among cities through wire transfers payments, banks may have a direct impact on how and when these cities engage either as suppliers or customers in economic transactions with other cities. This data is confidential and comes from the \textit{Sistema de Informações de Créditos} (SCR), which is maintained by the \textit{Banco Central do Brasil}.

    \item \textit{The inequality Gini index}: this variable controls for the inequality levels of each city. Inequality can have a deep impact on local economic activity and welfare. Therefore, they may reflect on how cities engage in new economic transactions with other cities. This data is public and comes from the population censuses of the IBGE.

    \item \textit{Dependence on external customers and suppliers}: this variable controls for the local economic structure of each municipality by introducing the level of dependence on suppliers and customers that are based outside the city. For instance, more dynamic and flexible economies may have less dependence with external suppliers and customers. These indicators are constructed using the STR, as previously defined.

    \item \textit{Concentration of the bank local credit portfolio}: this variable proxies for the concentration of bank credit to local firms in terms of their industry. We evaluate such measure using the Herfindahl-Hirschman Index (HHI) for city $i$ as follows:

    \begin{align}
        HHI^{\text{bank credit}}_{it} = \left(\frac{Agriculture_{it}}{Total \ Credit_{it}}\right)^2 + \left(\frac{Manufacturing_{it}}{Total \ Credit_{it}}\right)^2 + \left(\frac{Services_{it}}{Total \ Credit_{it}}\right)^2,
    \end{align}

    \noindent in which $Agriculture_{it}$, $Manufacturing_{it}$, and $Services_{it}$ are the outstanding credit that banks grant to firms in the agriculture, manufacturing, and services in city $i$ at year $t$. Moreover, $Total \ Credit_{it} = Agriculture_{it} + Manufacturing_{it} + Services_{it}$. This data is confidential and comes from the SCR.

    \item \textit{Concentration of the sectoral local jobs}: this variables controls for the concentration of job in different sectors of the city. We also compute this measure using HHI as follows:

    \begin{align}
        HHI^{\text{job}}_{it} &= \left(\frac{Manufacturing_{it}}{Total \ Jobs_{it}}\right)^2 + \left(\frac{Construction_{it}}{Total \ Jobs_{it}}\right)^2 + \left(\frac{Trade_{it}}{Total \ Jobs_{it}}\right)^2 + \nonumber\\
        & \qquad \left(\frac{Services_{it}}{Total \ Jobs_{it}}\right)^2 + \left(\frac{Agriculture_{it}}{Total \ Jobs_{it}}\right)^2,
    \end{align}

    \noindent in which $Manufacturing_{it}$, $Construction_{it}$, $Trade_{it}$, $Services_{it}$, $Agriculture_{it}$ are the total number of formal jobs in the manufacturing, construction, trade, services, and agriculture sectors, respectively, in city $i$ at year $t$. In addition, $Total \ Jobs_{it} = Manufacturing_{it} + Construction_{it}+ Trade_{it} + Services_{it} + Agriculture_{it}$. This data is public and comes from the \textit{Relação Anual de Informações Sociais} (RAIS), which is maintained by the \textit{Ministério da Economia}.

    \item \textit{Human development index (HDI)}: this variable controls for the human development levels in three different dimensions: income, health, and education. We take this data from Firjan, which is a network of private nonprofit organizations with more than ten thousand associates that computes the human development index for each city in Brazil since 2005.

    \item \textit{Courts efficiency}: this variable controls for the courts efficiency levels. This is a state-level variable. Therefore, we consider the same court efficiency level for all cities within the same state. More efficient courts may foster economic development because they reduce contractual frictions among firms. For instance, firms may be more willingful to provide trade credit to their firm counterparts because they know that they could execute their guarantees more rapidly in the local courts.

    To evaluate the court efficiency level, we use a  DEA model that optimally compares efficiency across different state-level courts. We measure efficiency using the output orientation, i.e., more efficient courts are those that output more with the same number of inputs. We use two inputs: (i) litigation backlog size, which quantifies the processes that are pending in the period, excluding those suspended, in a provisional file and for tax and criminal executions; and (ii) staff and judges expenditures, excluding expenses with retired staff. Our output variable is the number of complete cases. We follow the literature and use non-increasing returns to scale in our DEA model. We take these data from the \textit{Justiça Nacional} database, which is maintained by the \textit{Conselho Nacional de Justiça}.

\end{itemize}

Table \ref{tab:summary-stats} provides summary statistics of all variables employed in the econometric specification in \eqref{eq:spec}. We apply a Z-score transformation (subtract mean then divide by standard deviation) on all outcome variables employed in this paper. These variables end up with a zero mean and unit standard deviation. Therefore, our marginal effects are in terms of standard deviations relative to mean of the outcome variables.

\begin{table}[!htbp] \centering
  \caption{Summary statistics of the numerical variables employed in the econometric exercise.}
  \label{tab:summary-stats}
  \setlength\tabcolsep{3 pt}
  \footnotesize
\begin{tabular}{@{\extracolsep{5pt}}lccccccc}
\toprule\midrule
Statistic & \multicolumn{1}{c}{N} & \multicolumn{1}{c}{Mean} & \multicolumn{1}{c}{St. Dev.} & \multicolumn{1}{c}{Min} & \multicolumn{1}{c}{Pctl(25)} & \multicolumn{1}{c}{Pctl(75)} & \multicolumn{1}{c}{Max} \\
\midrule
\textit{Network measures}\\
Downstream centrality & 66,783 & 0.0004 & 0.006 & 0.00003 & 0.00003 & 0.00004 & 0.232 \\
Upstream centrality & 66,783 & 0.0002 & 0.003 & 0.00003 & 0.00003 & 0.0001 & 0.177 \\
Number of customers & 66,783 & 67.630 & 165.710 & 0 & 7 & 56 & 4,492 \\
Number of suppliers & 66,783 & 67.859 & 190.119 & 0 & 4 & 53 & 5,284 \\
Total received (R\$ bi) & 66,572 & 1.454 & 35.194 & 0.000 & 0.001 & 0.047 & 3,212.708 \\
Total paid (R\$ bi) & 63,690 & 1.605 & 39.423 & 0.000 & 0.001 & 0.026 & 3,343.833 \\
Dependence on External Customers  & 66,572 & 0.767 & 0.264 & 0.000 & 0.596 & 0.984 & 1.000 \\
Dependence on External Suppliers & 63,690 & 0.619 & 0.311 & 0.000 & 0.334 & 0.913 & 1.000 \\
\midrule
\textit{City characteristics}\\
GDP (R\$ bi) & 66,783 & 0.948 & 9.331 & 0.004 & 0.036 & 0.230 & 628.065 \\
Exports / GDP & 66,783 & 0.023 & 0.132 & 0.000 & 0.000 & 0.003 & 22.716 \\
Total credit / GDP & 66,783 & 0.010 & 0.075 & 0.00001 & 0.002 & 0.008 & 9.664 \\
Gini index & 66,781 & 0.514 & 0.067 & 0.099 & 0.472 & 0.558 & 0.808 \\
Bank financing concentration (HHI) & 66,783 & 0.636 & 0.188 & 0.333 & 0.501 & 0.760 & 1.000 \\
Sectoral jobs concentration (HHI) & 66,771 & 0.501 & 0.206 & 0.207 & 0.337 & 0.627 & 1.000 \\
Human development index (HDI) & 55,121 & 0.627 & 0.125 & 0.187 & 0.539 & 0.721 & 0.936 \\
Courts efficiency & 56,084 & 0.665 & 0.140 & 0.415 & 0.554 & 0.747 & 0.970\\
\midrule\bottomrule
\end{tabular}
\end{table}

Table \ref{tab:regression-results} reports coefficient estimates of Regression \ref{eq:spec} when we use the following network measures as outcomes: downstream centrality (Column (1)), upstream centrality (Column (2)), number of customers or in-degree (Column (3)), number of suppliers or out-degree (Column (4)), total received or in-strength (Column (5)), and total paid or out-strength (Column (6)).

If a city has a high downstream centrality, then it connects many customers with suppliers (which are the main providers of goods and services in the supply chain). This suggests that cities with high downstream centrality are vital in the economy. These cities are crucial in the development process. Our empirical results suggest a positive and statistically significant relationship between downstream centrality and the log of the city GDP (Column (1) of Table~\ref{tab:regression-results}), further corroborating the scatterplots in Figure \ref{fig:scat}. We find that a 1\% increase in GDP is correlated with an increase of 0.12 standard deviation of the downstream centrality. Therefore, high downstream centrality cities connect suppliers to the firms that are selling goods and services to final clients, both wholesale and retail. Therefore, a positive correlation is expected.

We also find a positive relationship between downstream centrality and the HHI of sectoral jobs. An increase of 1 percentage point in the HHI (sectoral jobs) is correlated with an increase of 0.632 in the standard deviation of the downstream centrality. Our results suggest that jobs specialization (concentration) may impact positively the downstream centrality of the city, giving support for the specialization theory of David Ricardo. The Gini coefficient is also statistically significant and positive (0.498), which suggests that there is a correlation between downstream centrality and income concentration.

Cities with high downstream centrality are cities that possess a large number of clients in the supply chain. These cities distribute resources to the rest of Brazil. They are therefore important due to their role in the supply chain. The other side of the coin would be cities with high upstream centrality. These cities receive a lot of resources from places that depend heavily on them. The buyer depends on the supply provided by firms in that city. Therefore, we can conceptualize the upstream centrality as a form of dependency of the city as a supplier in the entire supply chain.

In the Column (2) of Table~\ref{tab:regression-results}, we present results for the upstream centrality. The Gini coefficient has a negative correlation with the upstream centrality, that is, the higher the concentration of income, the lower the upstream centrality. As the upstream centrality connotes the city centrality from the supplier perspective to the entire supply chain, the more equitable in terms of income, the more relevant the city will be in this respect. Greater income distribution may allow more businesses to flourish that produce goods and services that are essential to other businesses, allowing for increased trade in the supply of goods and services.

The HHI for sectoral jobs also has a negative correlation, though marginally significant. This result suggests that, if the work is more concentrated in certain sectors, the city has less upstream centrality. As reflected by the Gini coefficient, the more concentrated the jobs in a sector are, the less capacity firms will have to provide a diverse list of goods and services, becoming relevant in the global production chain of goods and services.

It is interesting to note that higher courts efficiency associates with lower upstream centrality. This may be due to the fact that, in cities where Justice is more efficient, there may be more lawsuits and more litigation. In this specific case, for suppliers it may be worse to establish the base of their operations in these type of city.

There is a positive relationship between courts efficiency and the number of customers (Column (3)) and suppliers (Column (4)). This suggests that, as courts are more efficient in a specific city, the more firms engage in transactions with customers and suppliers from other cities, which reflects a positive impact of the Judiciary efficiency on economic transactions. This may be associated with the fact that more distant firms are relatively more opaque. Therefore, having a relative more efficient Judiciary system enables firms to engage in more distant cities, with new and potentially risky clients.  Brazil suffers from critics that the Judiciary is relatively inefficient, which hurts economic transactions. Our results put some light on this relationship and argues for policies that increases courts efficiency as an important mechanism that may reduce contractual frictions and therefore increase economic transactions and economic growth.

The Human Development Index (HDI) for income, health, and education, is negatively associated to the number of customers and suppliers. This result suggests that, as development indicators improve, the self-sufficiency of the city also increases.

Several factors help explain the total amount received (Column (5)) and total amount paid (Column (6)) for cities. The size of the city has a positive and statistical relationship with these two variables, which is intuitive as we would expect a size effect. The Gini coefficient is also positive, suggesting that higher inequality implies higher received and paid volumes.

Our findings also suggests that, as the dependence on external customers increase, the total amount received and paid  becomes smaller, whereas the effect of dependence of external suppliers is the opposite. Trade is higher between cities that have more dependence on external suppliers and is lower between cities that depend more on external customers.

\begin{table}[!htbp] \centering
  \caption{This table reports coefficient estimates of Regression \ref{eq:spec}. We apply a Z-score transformation in all outcome variables. Therefore, marginal effects are in terms of standard deviations of the outcome variables. Due to the region-time fixed effects, all coefficients are estimated by comparing municipalities within a same region. Errors are clustered at the city level. Statistical significance levels: $^{*}$p$<$0.1; $^{**}$p$<$0.05; $^{***}$p$<$0.01}
  \label{tab:regression-results}
  \setlength\tabcolsep{5 pt}
  \footnotesize
\begin{tabular}{@{\extracolsep{5pt}}lcccccc}
\toprule\midrule
\\[-1.8ex] \textit{Dependent variable:} & Downstream  & Upstream  & Number of & Number of & Total & Total  \\
\\[-3.8ex]  &  Centrality &  Centrality & Customers & Suppliers & Received & Paid  \\\cline{2-7}
\\[-1.8ex] & (1) & (2) & (3) & (4) & (5) & (6)\\
\hline \\[-1.8ex]
 log(GDP) & 0.120$^{**}$ & 0.002 & $-$0.00004 & 0.004 & 0.112$^{**}$ & 0.106$^{**}$ \\
  & (0.055) & (0.004) & (0.004) & (0.004) & (0.054) & (0.052) \\
[6pt]
 Exports/GDP & $-$0.279 & $-$0.038 & 0.007 & $-$0.015 & $-$0.239 & $-$0.240 \\
  & (0.238) & (0.026) & (0.024) & (0.019) & (0.198) & (0.201) \\
[6pt]
 Credit/GDP & $-$0.012 & $-$0.017 & $-$0.015 & $-$0.020 & $-$0.025 & $-$0.023 \\
  & (0.080) & (0.016) & (0.026) & (0.023) & (0.080) & (0.079) \\
[6pt]
 Gini & 0.498$^{*}$ & $-$0.291$^{**}$ & 0.193$^{*}$ & 0.113 & 0.540$^{*}$ & 0.535$^{*}$ \\
  & (0.290) & (0.115) & (0.110) & (0.119) & (0.323) & (0.320) \\
[6pt]
 Dependence on & $-$0.263$^{*}$ & $-$0.051 & $-$0.043 & $-$0.036 & $-$0.187$^{**}$ & $-$0.195$^{**}$ \\
 \ \ external customers & (0.137) & (0.050) & (0.036) & (0.036) & (0.093) & (0.096) \\
[6pt]
  Dependence on & 0.137$^{*}$ & $-$0.028 & 0.029 & 0.011 & 0.111$^{*}$ & 0.119$^{*}$ \\
   \ \ external suppliers & (0.076) & (0.025) & (0.021) & (0.021) & (0.067) & (0.069) \\
[6pt]
 HHI (bank financing) & $-$0.058 & 0.018 & 0.003 & 0.011 & 0.009 & 0.0002 \\
  & (0.053) & (0.039) & (0.037) & (0.039) & (0.032) & (0.031) \\
[6pt]
 HHI (sectoral jobs) & 0.632$^{*}$ & $-$0.057$^{*}$ & 0.014 & 0.038 & 0.538$^{*}$ & 0.528$^{*}$ \\
  & (0.322) & (0.033) & (0.044) & (0.046) & (0.275) & (0.275) \\
[6pt]
 HDI & $-$0.192 & 0.157$^{**}$ & $-$0.315$^{***}$ & $-$0.222$^{***}$ & $-$0.174 & $-$0.190 \\
  & (0.213) & (0.079) & (0.079) & (0.081) & (0.175) & (0.181) \\
[6pt]
 Courts efficiency & $-$0.137 & $-$0.320$^{***}$ & 0.827$^{***}$ & 0.518$^{***}$ & $-$0.096 & $-$0.098 \\
  & (0.184) & (0.103) & (0.086) & (0.089) & (0.150) & (0.154) \\
[6pt]
\midrule
\textit{Fixed effects}\\
Time $\cdot$ Region & Yes & Yes  & Yes  & Yes  & Yes  & Yes\\
\midrule
Observations & 53,481 & 53,481 & 53,481 & 53,481 & 53,481 & 53,481 \\
R$^{2}$ & 0.158 & 0.032 & 0.315 & 0.278 & 0.094 & 0.096 \\
Error clustering & City & City & City & City & City & City\\
\midrule\bottomrule
\end{tabular}
\end{table}

\section{Conclusion}
\label{sec:conclusions}

This paper uses a novel dataset composed of wire transfers to study how cities interconnect in a large and important emerging country. Brazil has over 5,500 municipalities, which together potentially create a very heterogeneous topological structure among different cities. The application of complex network theory in this type of economic network enables us to understand how supply chains interconnect across different cities, how central (important) are different cities, and how its structure changes over time.


We observe that the trade network has a pronounced disassortative mixing pattern, which relates to the power-law shape of firm distributions in Brazil (many small firms and few large firms). We also find that the assortativity drops after the Brazilian recession in 2014, which may be related to the fact that small firms failed and large firms concentrated even more economic flows.

We find a large degree of economic integration among cities, which we measure using the dependence on external suppliers and customers. This high coupling corroborates the high degree of specialization of cities either in agricultural, industry, or services activities. This evidence favors David Ricardo theory, in which cities should specialize in what they enjoy comparative advantage.

Municipal capitals are those that tend to connect cities that are very separate and, therefore, are the most likely candidates to be centers of the supply chain in Brazil. The relative importance of some cities varies over time. We predict that the ten main cities identified remain the same regardless of the centrality of the network and the year analyzed. We see more oscillations of the other cities.

Using the network density, we find that the trade network is very sparse, which confirms the existence of small urban centers that link very remote regional cities.

We find a network core consisting of cities in the southeast region (metropolitan areas of Belo Horizonte, Campinas, Rio de Janeiro and Sao Paulo), the southern region (metropolitan areas of Curitiba and Porto Alegre) and the Federal District.

Our results are important for the design of public policies as we can track down the most relevant cities in a connectedness perspective. Which cities perform a function of core in the network and may help speed up economic growth. Further research could explore the network using a variety of networks measures to deepen our understanding of the evolution of these networks.


\clearpage

\end{document}